\documentclass[10pt,pra,twocolumn,floatfix]{revtex4}
\usepackage{amsfonts}
\usepackage{amssymb}
\usepackage{amsmath}
\usepackage[dvips]{graphicx}
\usepackage{color}

\definecolor{red}{rgb}{1,0,0}

\begin{document}

\title{Topological error correction with a Gaussian cluster state}
\author{Shuhong Hao$^{1}$, Meihong Wang$^{2,3}$, Dong Wang$^{1}$, and
Xiaolong Su$^{2,3}$}
\email{suxl@sxu.edu.cn}
\affiliation{$^{1}$School of Mathematics and Physics, Anhui University of Technology,
Maanshan, 243000, People's Republic of China\\
$^{2}$State Key Laboratory of Quantum Optics and Quantum Optics Devices,
Institute of Opto-Electronics, Shanxi University, Taiyuan, 030006, People's
Republic of China\\
$^{3}$Collaborative Innovation Center of Extreme Optics, Shanxi University,
Taiyuan, Shanxi 030006, People's Republic of China}

\begin{abstract}
Topological error correction provides an effective method to correct errors
in quantum computation. It allows quantum computation to be implemented with
higher error threshold and high tolerating loss rates. We present a
topological a error correction scheme with continuous variables based on an
eight-partite Gaussian cluster state. We show that topological quantum
correlation between two modes can be protected against a single quadrature
phase displacement error occurring on any mode and some of two errors
occurring on two modes. More interestingly, some cases of errors occurring
on three modes can also be recognised and corrected, which is different from
the topological error correction with discrete variables. We show that the
final error rate after correction can be further reduced if the modes are
subjected to identical errors occurring on all modes with equal probability.
The presented results provide a feasible scheme for topological error
correction with continuous variables and it can be experimentally
demonstrated with a Gaussian cluster state.
\end{abstract}

\maketitle

\section{INTRODUCTION}

Quantum computation (QC) can solve many complex problems more efficiently
than classical computer \cite{Nielsen2000}. The measurement-based one-way QC
provides a practical model to perform the universal QC based on the cluster
states with different structures \cite{Raussendorf2001,Menicucci2006}.
Quantum logic gates are realized by measurement and feedforward of
measurement results based on the prepared cluster states in the
measurement-based one-way QC \cite{Raussendorf2001,Menicucci2006}. However,
during the QC, loss and noise may inevitably lead to errors in the
computation results. Many quantum error correction (QEC) schemes have been
experimentally demonstrated attempting to solve this problem \cite%
{Aoki2009,Lassen2010,Lassen2013,Hao}. However, most fault-tolerant QCs with
a high threshold error rate are difficult to implement in practice \cite%
{Knill,Aliferis}.

It has been shown that, cluster states whenever the underlying interaction
graph can be embedded in a three-dimensional cell structure, the so-called
cell complex \cite{Hatcher}, can be used for QEC and fault-tolerant QC \cite%
{Fowler,Raussendorf,Stern}. With a large cell complex, the quantum
algorithms can be realized by suitable braiding-like manipulation of the
defects \cite{Fowler}. It is based on the property that some topological
quantum correlations hold on defect-enclosing closed surfaces. We can use
the redundancy of the cell complex to protect the topological correlations
against local errors \cite{Raussendorf}. By using the topological properties
of the cluster states, we can realize the topological QC and the active
topological error correction (TEC) at the same time \cite%
{Fowler,Raussendorf,Stern}. This topological QC will lead to a higher error
threshold \cite{Wang,Raussendorf2007} and high tolerating loss rates \cite%
{Barrett} in the scalable QC.

So far, there are many experimental explorations about the topological
properties in a small topological quantum code unit. The anyonic fractional
statistics have been demonstrated in different physical systems, such as
photonic the system \cite{Chao-Yang,Liu}, the superconducting quantum
circuit \cite{Zhong,Song}, the ultracold atom system \cite{Han-Ning}, and
nuclear magnetic resonance systems \cite{Feng,Guilu}. The anyons can serve
as the fundamental units for a fault-tolerant QC. The TEC has been
experimentally demonstrated using a simpler eight-photon graph state \cite%
{Yao2012}, where the quantum correlation is protected against a single local
$Z$ error. This TEC method can significantly reduce the error rate.

Continuous variable (CV) QC, where information is encoded in the amplitude
and phase quadratures of photonic harmonic oscillators, can be realized
deterministically and unconditionally \cite{Menicucci2006,RMP1,RMP2}. The CV
cluster state is a basic resource for one-way CV QC \cite{Menicucci2006} and
quantum networks \cite{Wangyu,Su2020}. Recently, large-scale CV cluster
states in time the domain \cite{Andersen2019,Furusawa2019} and with
frequency comb \cite{Olivier2014,Cai2017} have been prepared experimentally,
and provide sufficient quantum resources for one-way CV QC. Several basic
quantum logic operations \cite{Wang2010,Ukai2011,Ukai20112}\ and even a gate
sequence \cite{Su2013} have been experimentally demonstrated in CV QC. What
is more, several feasible schemes for the quantum cubic phase gate have been
proposed \cite{Kevin,Kazunori}, which indicates that the full set of the
basic operations will soon be obtained.

In the regime of CV QEC, according to the no-go theorem for Gaussian QEC
that Gaussian errors cannot be corrected by using only Gaussian resources
\cite{Niset,Vuillot}, the linear oscillator codes \cite%
{Vuillot,Braunstein1998} are not suitable to correct generic Gaussian
errors, while the code introduced by Gottesman, Kitaev, and Preskill (GKP
code) \cite{Gottesman}, toric GKP code \cite{Vuillot} and the non-Gaussian
oscillator-into-oscillators code \cite{Noh2019} can correct generic Gaussian
errors. However, stochastic errors in CV QEC, which frequently occur in
channels with environment fluctuations for example, can easily cause
displacements and any other errors decomposable into displacements
(including non-Gaussian errors) \cite{Loock2010}. In the stochastic error
model, the input state described by the Wigner function $W_{in}$ is
transformed into a state $W_{error}$ with probability $\gamma $, and it
remains unchanged with probability $1-\gamma $ \cite{Loock2010}. Thus the
output state is given by ${W}_{out}(x,p)=(1-\gamma ){W}_{in}(x,p)+\gamma {W}%
_{error}(x,p)$. Note that even in the case that $W_{in}$ and $W_{error}$ are
two Gassian states, the output state $W_{out}$ is no longer Gaussian. Thus,
this error model describes a certain, simple form of non-Gaussian errors and
it can be corrected by Gaussian states and Gaussian operations. Some CV QEC
schemes against displacement errors have been experimentally demonstrated,
for example, the nine-wave-packet code \cite{Aoki2009}, the five-wave-packet
code \cite{Hao} and the correcting code with the correlated noisy channels
\cite{Lassen2013}. In addition, some basic concepts related to CV
topological QC have been proposed in recent years, such as the CV anyon
statistics \cite{Zhang2008}, the graphical calculus for CV states \cite%
{Menicucci2011}, the CV topological codes \cite{Tomoyuki} and its
application in quantum communication \cite{Menicucci2018}, the CV QC with
anyons \cite{Darran}, the exploration of CV fault-tolerant QC \cite%
{Menicucci2014,Lund}, and topological entanglement entropy \cite%
{Tomoyuki2014}. These works established the foundation for the further
research of CV topological QC. However, there has been no concrete scheme
for CV TEC up to now.

In this paper, we propose a concrete scheme for CV TEC based on an
eight-partite CV cluster state. At first, we propose the preparation scheme
for a topological eight-partite CV cluster state, which is obtained by
coupling eight squeezed states on a special beam-splitter network, and then
we present the CV TEC scheme. We show that, within the abilities of the
current technique, the quantum correlation can be protected against a single
quadrature phase displacement error occurring on any modes. Moreover, some
of the identical phase displacement errors occurring on two or three modes
at the same time can also be recognized and corrected. This shows that the
final error rate can be further reduced when we consider the phase sign of
the syndrome results. The presented results are an essential step in CV TEC
and are useful for further application in fault-tolerant CV QC.

The paper is organized as follows. In Sec. II, we present the preparation
scheme for the topological eight-partite CV cluster state. In Sec. III, we
show the details of the CV TEC scheme for a single mode error including the
details of error recognition and error correction procedures. In section IV,
we analyze the error rate of the presented CV TEC scheme. Finally, we
present the discussion and conclusion in Sec. V.

\section{THE TOPOLOGICAL CV CLUSTER STATE}

\begin{figure*}[tbp]
\centerline{
\includegraphics[width=160mm]{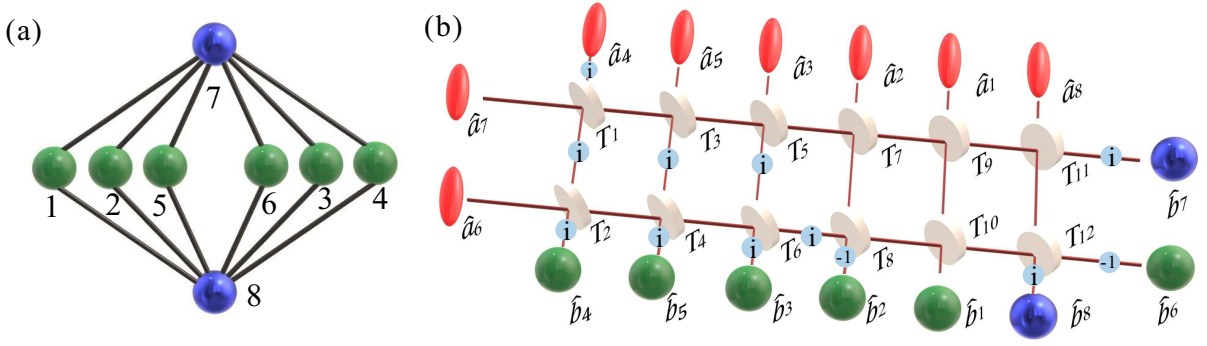}}
\caption{(a) The graph structure of the topological eight-partite CV cluster
state. The connected lines represent the interaction between the neighbor
nodes. Each node stands for an optical mode. (b) The beam-splitter network
for the preparation of the cluster state. The ellipsoids represent the input
squeezing states. The spheres represent the output cluster states. Circles
including $i$ are Fourier transforms and $-1$ is a $180^{\circ}$ rotation in
phase space.}
\end{figure*}

Similar to the discrete variable systems, the CV analog of the Pauli
operators are the Weyl--Heisenberg group of phase-space displacements. In
detail, the relationships between them are $X\rightarrow \hat{X}(t)=e^{-it%
\hat{p}}$ and $Z\rightarrow \hat{Z}(s)=e^{is\hat{x}}$, where the amplitude ($%
\hat{x}$) and phase ($\hat{p}$) quadratures of an optical mode $\hat{a}$\
are defined as $\hat{x}=(\hat{a}+\hat{a}^{\dagger })/2$\ and $\hat{p}=(\hat{a%
}-\hat{a}^{\dagger })/2i$. The eight-partite CV\ cluster state for the CV
TEC is described by a graph structure, as shown in the Fig. 1 (a). The CV
cluster state is defined as \cite{Zhang2006,Menicucci2011}
\begin{equation}
\hat{p}_{a}-\sum_{b\in N_{a}}\hat{x}_{b}\equiv \hat{\delta}_{a}\rightarrow
0,\ \ a\in G.
\end{equation}%
In the limit of infinite squeezing, the linear combinations of the
quadrature components (so-called nullifiers) in Eq. (1) tend to zero. The
modes $a\in G$\ denote the vertices of the graph $G$, while the modes $b\in
N_{a}$\ are the nearest neighbors of mode $\hat{a}$. The CV cluster state
can be generated with offline squeezing states and an appropriate
beam-splitter network \cite{vanLoock2007,Su2007,Mitsuyoshi,Su2012}.

The topological cluster state can be prepared by implementing an appropriate
unitary transformation $U$\ on a series of $\hat{p}$-squeezed input states, $%
\hat{a}_{l}=e^{+r}\hat{x}_{l}^{(0)}+ie^{-r}\hat{p}_{l}^{(0)}$, where $r$\ is
the squeezing parameter, $l=1,...,8,$and $\hat{x}^{(0)}$\ and $\hat{p}^{(0)}$%
\ represent the quadratures of a vacuum state whose variance is $%
\left\langle \Delta ^{2}\hat{x}^{(0)}\right\rangle =\left\langle \Delta ^{2}%
\hat{p}^{(0)}\right\rangle =1/4$, which corresponds to the shot-noise-level
(SNL). Then, the output modes can be obtained by $\hat{C}_{k}=\sum%
\nolimits_{l}U_{kl}\hat{a}_{l}$. According to the method of building a
Gaussian cluster state by linear optics \cite{vanLoock2007}, the
transformation matrix $U$\ should satisfy the condition $I$Im$U=A$Re$U$,
where $I$ is the identity matrix and $A$ is the adjacency matrix of the
graph $G$. Based on the unitarity of matrix $U$, we obtain Re$U($Re$%
U)^{T}=(I+A^{2})^{-1}$. We need $n(n-1)/2$ auxiliary conditions to get the
matrix $U$. To make the matrix $U$ simple, we choose the auxiliary
conditions to make more elements of $U$ to be zero according to symmetry of $%
A$. Finally, we get a simple form of $U$\ as follows
\begin{equation}
\left(
\begin{array}{cccccccc}
\sqrt{\frac{3}{5}} & \frac{2}{\sqrt{35}} & \frac{2}{3\sqrt{7}} & \frac{-i}{%
\sqrt{91}} & \frac{2}{3\sqrt{11}} & \frac{i}{\sqrt{7}} & \frac{2}{\sqrt{143}}
& 0 \\
0 & \frac{-5}{\sqrt{35}} & \frac{2}{3\sqrt{7}} & \frac{-i}{\sqrt{91}} &
\frac{2}{3\sqrt{11}} & \frac{i}{\sqrt{7}} & \frac{2}{\sqrt{143}} & 0 \\
0 & 0 & \frac{-\sqrt{7}}{3} & \frac{-i}{\sqrt{91}} & \frac{2}{3\sqrt{11}} &
\frac{i}{\sqrt{7}} & \frac{2}{\sqrt{143}} & 0 \\
0 & 0 & 0 & \frac{-i}{\sqrt{91}} & 0 & \frac{i}{\sqrt{7}} & \frac{-11}{\sqrt{%
143}} & 0 \\
0 & 0 & 0 & \frac{-i}{\sqrt{91}} & \frac{-3}{\sqrt{11}} & \frac{i}{\sqrt{7}}
& \frac{2}{\sqrt{143}} & 0 \\
\frac{-2}{\sqrt{15}} & \frac{2}{\sqrt{35}} & \frac{2}{3\sqrt{7}} & \frac{-i}{%
\sqrt{91}} & \frac{2}{3\sqrt{11}} & \frac{i}{\sqrt{7}} & \frac{2}{\sqrt{143}}
& \frac{1}{\sqrt{3}} \\
\frac{i}{\sqrt{15}} & \frac{-i}{\sqrt{35}} & \frac{-i}{3\sqrt{7}} & \frac{-7%
}{\sqrt{91}} & \frac{-i}{3\sqrt{11}} & 0 & \frac{-i}{\sqrt{143}} & \frac{i}{%
\sqrt{3}} \\
\frac{i}{\sqrt{15}} & \frac{-i}{\sqrt{35}} & \frac{-i}{3\sqrt{7}} & \frac{6}{%
\sqrt{91}} & \frac{-i}{3\sqrt{11}} & \frac{1}{\sqrt{7}} & \frac{-i}{\sqrt{143%
}} & \frac{i}{\sqrt{3}}%
\end{array}%
\right) .
\end{equation}

Generally, we can decompose an arbitrary $n\times n$ matrix $U$ into an at
most $n(n+1)/2$ beam-splitter network \cite{Michael}. Here, we decompose it
symmetrically in order to use beam-splitters as little as possible. The
decomposed matrix is $U=I_{2}(-1)\ F_{3}\ F_{4}$ $F_{5}$ $I_{6}(-1)$ $F_{7}$
$F_{8}$ $B_{68}^{+}(T_{12})$ $B_{78}^{+}(T_{11})$ $B_{16}^{+}(T_{10})$ $%
B_{17}^{+}(T_{9})$ $B_{26}^{+}(T_{8})$ $F_{6}$ $B_{27}^{+}(T_{7})$ $%
B_{36}^{+}(T_{6})$ $F_{3}$ $B_{37}^{-}(T_{5})$ $B_{56}^{-}(T_{4})$ $F_{5}$ $%
B_{57}^{-}(T_{3})$ $B_{46}^{-}(T_{2})$ $F_{4}$ $B_{47}^{+}(T_{1})$ $F_{4}$.
In detail, $F_{k}$\ denotes the Fourier transformation of mode $k$, which
corresponds to a 90$^{\circ }$\ rotation in the phase space; $B_{kl}^{\pm
}(T_{j})$\ stands for the linearly optical transformation on the $j$-th
beam-splitter with the transmittance of $T_{j}$\ ($j=1,2,\ldots 12$), where $%
(B_{kl}^{\pm })_{kk}=\sqrt{T}$, $(B_{kl}^{\pm })_{kl}=\sqrt{1-T}$, $%
(B_{kl}^{\pm })_{lk}=\pm \sqrt{1-T}$, and $(B_{kl}^{\pm })_{ll}=\mp \sqrt{T}$%
\ are elements of the beam-splitter matrix; and $I_{k}(-1)=e^{i\pi }$\
corresponds to a 180$^{\circ }$\ rotation of mode $k$\ in the phase space.
The transmittances of the 12 beam-splitters are chosen as $%
T_{1}=1/78,T_{2}=6/7,T_{3}=54/55,T_{4}=5/6,T_{5}=35/36,T_{6}=4/5,T_{7}=20/21,T_{8}=3/4,T_{9}=9/10,T_{10}=2/3,T_{11}=2/3,
$ and $T_{12}=1/2$, respectively. The beam-splitter network is shown in Fig.
1 (b).

The eight output modes $\hat{C}_{k}$ ($k=1,2,\ldots 8$) constitute the
topological eight-partite cluster state. For the input states with
finite-squeezing, the nullifiers of the cluster state are
\begin{equation}
\begin{array}{l}
\hat{\delta}_{1}=\hat{p}_{1}-(\hat{x}_{7}+\hat{x}_{8})=e^{-r}\frac{\sqrt{5}%
\hat{p}_{1}^{(0)}+2\hat{p}_{8}^{(0)}}{\sqrt{3}}, \\
\delta _{2}=\hat{p}_{2}-(\hat{x}_{7}+\hat{x}_{8})=e^{-r}(\frac{2\hat{p}%
_{1}^{(0)}}{\sqrt{15}}-\sqrt{\frac{7}{5}}\hat{p}_{2}^{(0)}+\frac{2\hat{p}%
_{8}^{(0)}}{\sqrt{3}}), \\
\hat{\delta}_{3}=\hat{p}_{3}-(\hat{x}_{7}+\hat{x}_{8})=e^{-r}(\frac{2\hat{p}%
_{1}^{(0)}}{\sqrt{15}}-\frac{2\hat{p}_{2}^{(0)}}{\sqrt{35}}-\frac{3\hat{p}%
_{3}^{(0)}}{\sqrt{7}}+\frac{2\hat{p}_{8}^{(0)}}{\sqrt{3}}), \\
\hat{\delta}_{4}=\hat{p}_{4}-(\hat{x}_{7}+\hat{x}_{8}) \\
\ \ \ =e^{-r}(\frac{2\hat{p}_{1}^{(0)}}{\sqrt{15}}-\frac{2\hat{p}_{2}^{(0)}}{%
\sqrt{35}}-\frac{2\hat{p}_{3}^{(0)}}{3\sqrt{7}}-\frac{2\hat{p}_{5}^{(0)}}{3%
\sqrt{11}}-\frac{\sqrt{13}\hat{p}_{7}^{(0)}}{\sqrt{11}}+\frac{2\hat{p}%
_{8}^{(0)}}{\sqrt{3}}), \\
\hat{\delta}_{5}=\hat{p}_{5}-(\hat{x}_{7}+\hat{x}_{8}) \\
\ \ \ =e^{-r}(\frac{2\hat{p}_{1}^{(0)}}{\sqrt{15}}-\frac{2\hat{p}_{2}^{(0)}}{%
\sqrt{35}}-\frac{2\hat{p}_{3}^{(0)}}{3\sqrt{7}}-\frac{\sqrt{11}\hat{p}%
_{5}^{(0)}}{3}+\frac{2\hat{p}_{8}^{(0)}}{\sqrt{3}}), \\
\hat{\delta}_{6}=\hat{p}_{6}-(\hat{x}_{7}+\hat{x}_{8})=\sqrt{3}e^{-r}\hat{p}%
_{8}^{(0)}, \\
\hat{\delta}_{7}=\hat{p}_{7}-(\hat{x}_{1}+\hat{x}_{2}+\hat{x}_{3}+\hat{x}%
_{4}+\hat{x}_{5}+\hat{x}_{6}) \\
\ \ \ =e^{-r}\frac{6\hat{p}_{6}^{(0)}-\sqrt{13}\hat{p}_{4}^{(0)}}{\sqrt{7}},
\\
\hat{\delta}_{8}=\hat{p}_{8}-(\hat{x}_{1}+\hat{x}_{2}+\hat{x}_{3}+\hat{x}%
_{4}+\hat{x}_{5}+\hat{x}_{6})=\sqrt{7}e^{-r}\hat{p}_{6}^{(0)},%
\end{array}%
\end{equation}%
respectively. In the case of infinite squeezing ($r\rightarrow \infty $),
these nullifiers trend to zero, which satisfies the definition of the CV
cluster state in Eq. (1).

\section{TOPOLOGICAL ERROR CORRECTION}

For a given cluster state, its nullifiers can be used as the generators of
the stabilizer operators for a topological code (a stabilizer QEC code) \cite%
{Tomoyuki}. The topological quantum correlations in the qubit system are
defined as $\mathbf{C}_{F}\equiv \langle \otimes _{f\in F}X_{f}\rangle =1$
\cite{Yao2012}. According to the corresponding relationship between the
qubit Pauli operation with discrete variables and the single-mode Weyl$-$%
Heisenberg operation in the CV system, the single-mode Weyl$-$Heisenberg
operator is described as $\hat{X}_{f}(t_{f})=e^{-it_{f}\hat{p}%
_{f}}(f=1,2,....,6)$ in the CV system, where~$t_{f}$~is equal to~$(-1)^{f}t$
\cite{Zhang2008}. By substituting the single-mode Weyl$-$Heisenberg operator
into the topological quantum correlation in the qubit system, the definition
of the topological quantum correlation in the CV system becomes $\mathbf{C}%
_{F}^{CV}\equiv \langle \otimes _{f\in F}\hat{X}_{f}[(-1)^{f}t]\rangle
=\langle \otimes _{f\in F}e^{-i(-1)^{f}t\hat{p}_{f}}\rangle =1$. In CV
system, it is convenient to use the nullifiers to analyze the protected
quantum correlations when displacement errors occurred. So we have $\langle
\sum_{f\in F}(-1)^{f}\hat{p}_{f}\rangle =0$ for the topological quantum
correlations in the CV cluster state. For the prepared eight-partite the CV
cluster state, the topological quantum correlations that can be protected
are
\begin{equation}
\begin{array}{l}
\hat{p}_{1}-\hat{p}_{2}=e^{-r}(\sqrt{\frac{3}{5}}\hat{p}_{1}^{(0)}+\sqrt{%
\frac{7}{5}}\hat{p}_{2}^{(0)}), \\
\hat{p}_{2}-\hat{p}_{5}=\frac{e^{-r}}{21}(-3\sqrt{35}\hat{p}_{2}^{(0)}+2%
\sqrt{7}\hat{p}_{3}^{(0)}+7\sqrt{11}\hat{p}_{5}^{(0)}), \\
\hat{p}_{3}-\hat{p}_{6}=e^{-r}(\frac{2}{\sqrt{15}}\hat{p}_{1}^{(0)}-\frac{2}{%
\sqrt{35}}\hat{p}_{2}^{(0)}-\frac{3}{\sqrt{7}}\hat{p}_{3}^{(0)}-\frac{1}{%
\sqrt{3}}\hat{p}_{8}^{(0)}), \\
\hat{p}_{4}-\hat{p}_{3}=\frac{e^{-r}}{33}(11\sqrt{7}\hat{p}_{3}^{(0)}-2\sqrt{%
11}\hat{p}_{5}^{(0)}-3\sqrt{143}\hat{p}_{7}^{(0)}), \\
\hat{p}_{5}-\hat{p}_{6}=e^{-r}(\frac{2\hat{p}_{1}^{(0)}}{\sqrt{15}}-\frac{2%
\hat{p}_{2}^{(0)}}{\sqrt{35}}-\frac{2\sqrt{7}\hat{p}_{3}^{(0)}}{21}-\frac{7%
\sqrt{11}\hat{p}_{5}^{(0)}}{21}-\frac{7\sqrt{3}\hat{p}_{8}^{(0)}}{21}),%
\end{array}%
\end{equation}%
respectively. The corresponding variances of the topological quantum
correalations are given by
\begin{equation}
\begin{array}{l}
\left\langle \Delta ^{2}(\hat{p}_{1}-\hat{p}_{2})\right\rangle =\frac{1}{2}%
e^{-2r}, \\
\left\langle \Delta ^{2}(\hat{p}_{2}-\hat{p}_{5})\right\rangle =\frac{1}{2}%
e^{-2r}, \\
\left\langle \Delta ^{2}(\hat{p}_{3}-\hat{p}_{6})\right\rangle =\frac{1}{2}%
e^{-2r}, \\
\left\langle \Delta ^{2}(\hat{p}_{4}-\hat{p}_{3})\right\rangle =\frac{1}{2}%
e^{-2r}, \\
\left\langle \Delta ^{2}(\hat{p}_{5}-\hat{p}_{6})\right\rangle =\frac{1}{2}%
e^{-2r},%
\end{array}%
\end{equation}%
respectively. Obviously, in the ideal case of infinite squeezing ($%
r\rightarrow \infty $), these excess noises will vanish and the better the
squeezing, the smaller the noise terms are.

At first, we analyze the CV TEC for a single phase displacement error in the
ideal case, i.e. in the case of infinite squeezing. In the error correction
procedure, any one of the above topological quantum correlations can be
protected with the other four "redundant" topological correlations as error
syndromes in the TEC. In the process of error recognition, we choose any one
of these five topological quantum correlations as the one needed to be
protected and the other four as the auxiliary quantum correlations to get
the error syndrome. Here, we take the quantum correlation $\hat{p}_{5}-\hat{p%
}_{6}$ as an example to analyze the TEC process. Figure 2 shows the circuit
for error syndrome recognition in the case of an error that occurred on the
mode $\hat{C}_{5}$. We use the homodyne detection systems to measure the
quantum correlation of phase quadratures of the optical modes $\hat{C}_{1}$
to $\hat{C}_{6}$, respectively. The circuits for measuring the error
syndrome can be realized by controlling the phase difference between the
local light and the measured mode to get the phase quadrature ($\hat{p}$)
and making an appropriate combination of the measured phase quadratures
according to Eq. (4) to obtain the topological correlations of $\hat{p}_{1}-%
\hat{p}_{2}$, $\hat{p}_{2}-\hat{p}_{5}$, $\hat{p}_{3}-\hat{p}_{6}$, and $%
\hat{p}_{4}-\hat{p}_{3}$, respectively. If an error occurs on any mode, the
quantum correlations that contain this mode will be affected at the same
time. For example, when an error occurs on mode $\hat{C}_{5}$, the
topological quantum correlation of $\hat{p}_{2}-\hat{p}_{5}$ will not be
zero anymore. So, we can locate the position of the error based on the error
syndromes of the auxiliary quantum correlations. Then, by feedforward of the
measurement results of $\hat{p}_{2}-\hat{p}_{5}$ to $\hat{p}_{5}-\hat{p}_{6}$%
, the effect of error on $\hat{p}_{5}-\hat{p}_{6}$ will be corrected.

Table 1 shows the error syndrome results for all kinds of different single
errors on modes $\hat{C}_{1}$ to $\hat{C}_{6}$ in the ideal case. If the
error $\varepsilon $ doesn't affect the syndrome correlation, the measured
correlation will remain unchanged. Otherwise, the measured syndrome
correlations will contain the nonzero error signal $\varepsilon $. The error
syndrome results are different from each other when the error occurs on
different modes. Comparing the measured results for the corresponding
quantum correlations $\hat{p}_{1}-\hat{p}_{2},$ $\hat{p}_{2}-\hat{p}_{5},$ $%
\hat{p}_{3}-\hat{p}_{6},$and $\hat{p}_{4}-\hat{p}_{3}$ with the predictions
in Table~I, we can identify the position of error.

\begin{table}[tbp]
\begin{tabular}{lllllc}
\multicolumn{6}{l}{TABLE I.\ \ Error syndrome with a single error mode.} \\
\hline\hline
Error mode\  & $\hat{p}_{1}-\hat{p}_{2}$ & $\hat{p}_{2}-\hat{p}_{5}$ & $\hat{%
p}_{3}-\hat{p}_{6}$ & $\hat{p}_{4}-\hat{p}_{3}$ & Requirement \\ \hline
\multicolumn{1}{c}{1} & \multicolumn{1}{c}{$\varepsilon $} &
\multicolumn{1}{c}{$0$} & \multicolumn{1}{c}{$0$} & \multicolumn{1}{c}{$0$}
& N \\
\multicolumn{1}{c}{2} & \multicolumn{1}{c}{$-\varepsilon $} &
\multicolumn{1}{c}{$\varepsilon $} & \multicolumn{1}{c}{$0$} &
\multicolumn{1}{c}{$0$} & N \\
\multicolumn{1}{c}{3} & \multicolumn{1}{c}{$0$} & \multicolumn{1}{c}{$0$} &
\multicolumn{1}{c}{$\varepsilon $} & \multicolumn{1}{c}{$-\varepsilon $} & N
\\
\multicolumn{1}{c}{4} & \multicolumn{1}{c}{$0$} & \multicolumn{1}{c}{$0$} &
\multicolumn{1}{c}{$0$} & \multicolumn{1}{c}{$\varepsilon $} & N \\
\multicolumn{1}{c}{5} & \multicolumn{1}{c}{$0$} & \multicolumn{1}{c}{$%
-\varepsilon $} & \multicolumn{1}{c}{$0$} & \multicolumn{1}{c}{$0$} & Y \\
\multicolumn{1}{c}{6} & \multicolumn{1}{c}{$0$} & \multicolumn{1}{c}{$0$} &
\multicolumn{1}{c}{$-\varepsilon $} & \multicolumn{1}{c}{$0$} & Y \\
\hline\hline
\end{tabular}%
\end{table}

After the position of error is confirmed, we can correct the error to
eliminate the affection of error on the protected correlation according to
the requirement. The correction requirements are summarized in Table I,
where the symbol N stands for no need to correct the error and\ the symbol Y
stands for the cases that need to correct the error. When the error occurs
on the modes $\hat{C}_{1}$, $\hat{C}_{2},$ $\hat{C}_{3},$ and $\hat{C}_{4}$
respectively, the protected quantum correlation $\hat{p}_{5}-\hat{p}_{6}$
will not be affected by the error. So we do not need to correct the error.
When an error $\varepsilon $ occurs on the mode $\hat{C}_{5},$ the
correlation becomes $\hat{p}_{5}-\hat{p}_{6}\rightarrow \varepsilon $. To
protect the correlation, we can make an addition of the detected term $\hat{p%
}_{2}-\hat{p}_{5},$ which tends to $-\varepsilon $, with $\hat{p}_{5}-\hat{p}%
_{6}$. We have $\hat{p}_{5}-\hat{p}_{6}+(\hat{p}_{2}-\hat{p}_{5})\rightarrow
0$ in the ideal case. So, the error influence is eliminated in this way. The
error for mode $\hat{C}_{6}$ can be corrected similarly by subtracting the
detected term $\hat{p}_{3}-\hat{p}_{6}$ from $\hat{p}_{5}-\hat{p}_{6}$. We
have $\hat{p}_{5}-\hat{p}_{6}-(\hat{p}_{3}-\hat{p}_{6})\rightarrow 0.$

\begin{figure}[tbp]
\centerline{
\includegraphics[width=75mm]{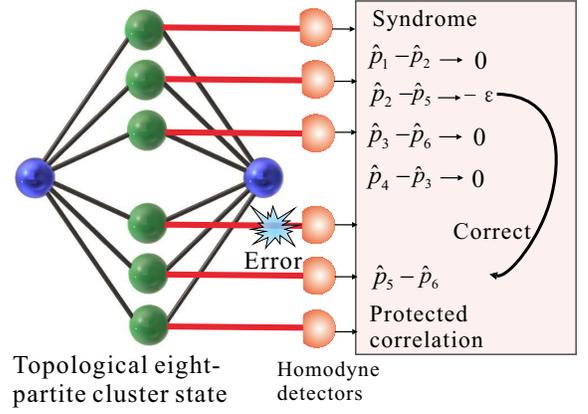}}
\caption{The circuit for error syndrome recognition and correction for the
CV TEC. Here, we take the case of an error that occurs on mode $\hat{C}_{5}$
as an example.}
\end{figure}

\begin{figure}[tbp]
\centerline{
\includegraphics[width=75mm]{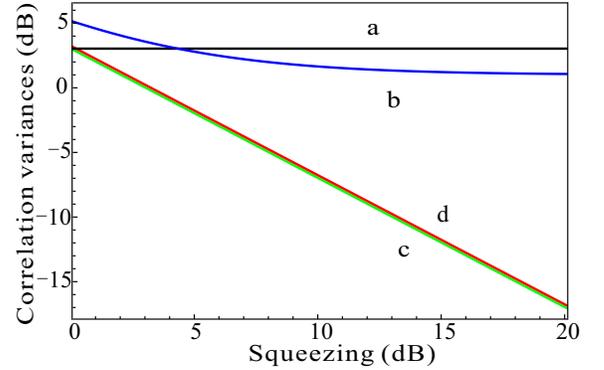}
}
\caption{The dependence of the quantum correlation of $\hat{p}_{5}-\hat{p}%
_{6}$ on the squeezing of the initial squeezed states for different cases.%
\textbf{\ }Curve a is the corresponding shot-noise level for the protected
correlation. Curve b is the correlation variance when no correction is used.
Curve c is the correlation variance when no error has occurred. Curve d is
the correlation variance when the TEC is used.}
\end{figure}

Then, we analyze the TEC in the case of real experimental parameters. We
consider an error $\varepsilon $ with a variance of $\varepsilon ^{2}=0.315$%
, which is 1 dB higher than the SNL, that occurred on the phase quadrature
of an optical mode. When the error occurs on mode $\hat{C}_{5}$, the
dependence of the protected topological quantum correlation $\hat{p}_{5}-%
\hat{p}_{6}$ on the squeezing of the initial squeezed states for different
cases is shown in Fig. 3. The correction effect is obvious when compared
with the curves with and without correction, i.e. curves b and d in Fig. 3.
To correct the error, the measurement results of $\hat{p}_{2}-\hat{p}_{5}$
is fedforward to $\hat{p}_{5}-\hat{p}_{6}$, and we have $\hat{p}_{5}-\hat{p}%
_{6}+(\hat{p}_{2}-\hat{p}_{5})=\hat{p}_{2}-\hat{p}_{6}=e^{-r}(\frac{2\hat{p}%
_{1}^{(0)}}{\sqrt{15}}-\frac{\sqrt{7}\hat{p}_{2}^{(0)}}{\sqrt{5}}-\frac{\hat{%
p}_{8}^{(0)}}{\sqrt{3}})$, whose variance is $\left\langle \Delta ^{2}(\hat{p%
}_{2}-\hat{p}_{6})\right\rangle =\frac{1}{2}e^{-2r}$, which is the same as
the variance of $\left\langle \Delta ^{2}(\hat{p}_{5}-\hat{p}%
_{6})\right\rangle =\frac{1}{2}e^{-2r}$ when there is no error. Thus, the
corrected correlation variance (curve d) is exactly the same with the
quantum correction when no error has occurred (curve c). This means that
using this TEC method, the error can not only be located but can also be
eliminated absolutely.

In the experiment, the photon loss is inevitable. However, the loss in the
whole experimental setup can be estimated and the noise will make the
variance of topological quantum correlation higher than that with loss. In
this case, we can evaluate the effect of loss on the topological quantum
correlation first. Then, we compare the measured variance of topological
quantum correlation with that with loss. If they are the same, it means that
there is no error. If the measured variance of topological quantum
correlation is higher than that with loss, it indicates the existence of
error caused by noise. Finally, we implement the corresponding TEC procedure
to remove errors.

\section{ERROR RATE OF THE CV TEC}

We show that the CV TEC scheme works for the case of single phase
displacement error, which usually occurs in the Markovian environment \cite%
{Aoki2009,Hao}, in the above section. However, the situation becomes more
complex when more than one error that occurs simultaneously. In this case,
the presented TEC scheme protects displacement errors coming from certain
noise which has special symmetry properties, such as identical errors for
different channels. Here, we consider the CV TEC scheme in the case of
identical errors occur simultaneously, which usually occurs in a
non-Markovian environment \cite{Lassen2013}. In practice, the noise in
different quantum channels exhibits correlations in time and space \cite%
{Lassen2013}. Thus it is necessary to consider quantum channels with a
correlated noise (non-Markovian environment), which corresponds to the case
that all optical modes are subjected to identical $\hat{p}$-displacement
error $\varepsilon $ with an equal probability $p$ simultaneously. When
errors occur on the modes $\hat{C}_{5}$ and $\hat{C}_{6}$ at the same time
with probability $p$, the quantum correlation $\hat{p}_{5}-\hat{p}_{6}$ will
not be influenced because the identical errors are canceled. This shows the
robustness of the topological correlation against two identical errors on
two optical modes. So the error rate for the protected correlation without
TEC is $P_{1}=1-p^{2}-(1-p)^{2}$. Actually, all the topological quantum
correlations $\hat{p}_{1}-\hat{p}_{2},$ $\hat{p}_{2}-\hat{p}_{5},$ $\hat{p}%
_{3}-\hat{p}_{6},$ $\hat{p}_{4}-\hat{p}_{3},$ and $\hat{p}_{5}-\hat{p}_{6},$
are robust to the case of errors that occurred on all six modes $\hat{C}_{1}$
to $\hat{C}_{6}$ at the same time.

There are several possibilities for the case of more than one error occurred
simultaneously with the same amplitude and probability. When identical
errors occurred on all six modes, we do not need to correct the errors as
discussed in the above paragraph, which corresponds to the case of zero
error. When there are five identical errors that have occurred on five
modes, this corresponds to the case that one error occurred on one mode,
which can be corrected in the way presented in Sec. III. The situation for
five identical errors has a similar syndrome results as that in Table I. We
can get the syndrome measurements table for the five identical errors just
by changing "error mode" into "mode without error" in Table I. The
correction method is exactly the same as that of one error. The
probabilities of one error and five identical errors are $6p^{5}(1-p)$ and $%
6p(1-p)^{5}$, respectively.

When there are four identical errors that have occurred on four modes, it
corresponds to the case that there are two identical errors that have
occurred on two modes simultaneously. Thus we only need to analyze the cases
for two and three identical\ errors that have occurred on two and three
modes simultaneously, respectively. The error syndrome measurements with two
identical errors are shown in Table II. Please note that the error modes
with and without brackets correspond to the syndrome correlations with and
without brackets, respectively. The cases of the first six lines can be
distinguished from each other and their syndrome measurements are also
different from all the cases in the Table I. So the error in these cases can
be recognized and corrected. The last three lines have syndrome measurements
identical to some lines in Table I. Unfortunately, the correction
requirements are different absolutely. So we cannot distinguish them (lines
7, 8 and 9 in Table II and one error on modes $\hat{C}_{5}$, $\hat{C}_{1}$,
and $\hat{C}_{2}$ in Table I, respectively) and make a right correction.
Generally, the error rate $p$ is always small. When the case that errors can
not be distinguished happens, the probability of two identical\ errors is
smaller than that of one error. We can select to correct the case of one
error in Table I to reduce the final error rate. A similar table for the
situation of four identical error modes can be obtained correspondingly. The
correction probabilities of two and four identical errors are $%
9p^{4}(1-p)^{2}$ and $9p^{2}(1-p)^{4},$ respectively.
\begin{table}[tbp]
\begin{tabular}{cccccc}
\multicolumn{6}{l}{TABLE II. \ Error syndrome with two identical errors.} \\
\hline\hline
The\ error & $\hat{p}_{1}-\hat{p}_{2}$ & $\hat{p}_{2}-\hat{p}_{5}$ & $\hat{p}%
_{3}-\hat{p}_{6}$ & $\hat{p}_{4}-\hat{p}_{3}$ & Require- \\
modes & \multicolumn{1}{l}{$(\hat{p}_{4}-\hat{p}_{3})$} & \multicolumn{1}{l}{%
$(\hat{p}_{3}-\hat{p}_{6})$} & \multicolumn{1}{l}{$(\hat{p}_{2}-\hat{p}_{5})$%
} & \multicolumn{1}{l}{$(\hat{p}_{1}-\hat{p}_{2})$} & ment \\ \hline
1,6 (4,5) & $\varepsilon $ & $0$ & $-\varepsilon $ & $0$ & Y \\
2,6 (3,5) & $-\varepsilon $ & $\varepsilon $ & $-\varepsilon $ & $0$ & Y \\
1,4 & $\varepsilon $ & $0$ & $0$ & $\varepsilon $ & N \\
5,6 & $0$ & $-\varepsilon $ & $-\varepsilon $ & $0$ & N \\
1,3 (2,4) & $\varepsilon $ & $0$ & $\varepsilon $ & $-\varepsilon $ & N \\
2,3 & $-\varepsilon $ & $\varepsilon $ & $\varepsilon $ & $-\varepsilon $ & N
\\ \hline
1,2 (4,3) & $0$ & $\varepsilon $ & $0$ & $0$ & N \\
2,5 (3,6) & $-\varepsilon $ & $0$ & $0$ & $0$ & Y \\
1,5 (4,6) & $\varepsilon $ & $-\varepsilon $ & $0$ & $0$ & Y \\ \hline\hline
\end{tabular}%
\end{table}

\begin{table}[tbp]
\begin{tabular}{cccccc}
\multicolumn{6}{l}{TABLE III.\ \ Error syndrome with three identical errors.}
\\ \hline\hline
The error & $\hat{p}_{1}-\hat{p}_{2}$ & $\hat{p}_{2}-\hat{p}_{5}$ & $\hat{p}%
_{3}-\hat{p}_{6}$ & $\hat{p}_{4}-\hat{p}_{3}$ & Require- \\
modes & $(\hat{p}_{4}-\hat{p}_{3})$ & $(\hat{p}_{3}-\hat{p}_{6})$ & $(\hat{p}%
_{2}-\hat{p}_{5})$ & $(\hat{p}_{1}-\hat{p}_{2})$ & ment \\ \hline
1,3,4(1,2,4) & $\varepsilon $ & $0$ & $\varepsilon $ & $0$ & N \\
2,3,4(1,2,3) & $-\varepsilon $ & $\varepsilon $ & $\varepsilon $ & $0$ & N
\\
2,4,5 & $-\varepsilon $ & $0$ & $0$ & $\varepsilon $ & Y \\
3,4,5 & $0$ & $-\varepsilon $ & $\varepsilon $ & $0$ & Y \\
1,4,6(1,4,5) & $\varepsilon $ & $0$ & $-\varepsilon $ & $\varepsilon $ & Y
\\
1,3,5 & $\varepsilon $ & $-\varepsilon $ & $\varepsilon $ & $-\varepsilon $
& Y \\ \hline
5,1,2 & $0$ & $0$ & $0$ & $0$ & Y \\ \hline\hline
\end{tabular}%
\end{table}

\begin{figure}[tbp]
\centerline{
\includegraphics[width=80mm]{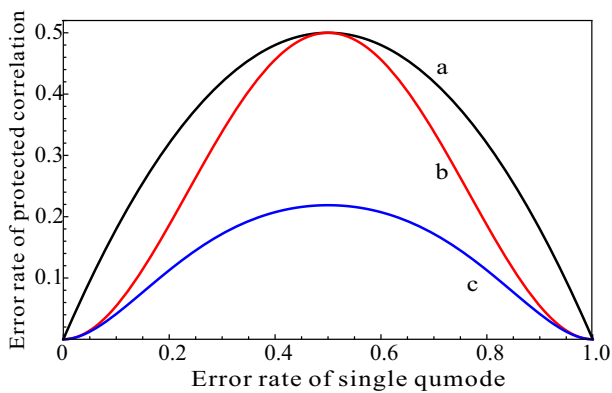}
}
\caption{The dependence of the error rates for the protected quantum
correlation on the single mode error rate $p$. Curve a isthe error rate for
the protected correlation without TEC. Curve b is the error rate of TEC for
qubit. Curve c is the error rate of the CV TEC.}
\end{figure}

The TEC with discrete variables presented in Ref. \cite{Yao2012} cannot deal
with the case of three identical errors. However, in the CV system, some of
these cases for three identical errors can be recognized and corrected. The
error syndrome measurements with three identical error modes are shown in
Table III. To distinguish them with the situations in Table II, we can
compare these nonzero syndrome measurements. For example, the cases of
errors that occurred on modes $\hat{C}_{1}$, $\hat{C}_{3}$, and $\hat{C}_{4}$
simultaneously in the first line of Table III and two errors that occurred
on modes $\hat{C}_{1}$ and $\hat{C}_{6}$ simultaneously in Table II have the
same nonzero syndromes, but the syndrome values of $\hat{p}_{1}-\hat{p}_{2}$
and $\hat{p}_{3}-\hat{p}_{6}$ are in-phase for the former and out-of-phase
for the latter. So, we can identify the case of three identical errors on
modes $\hat{C}_{1}$, $\hat{C}_{3}$, and $\hat{C}_{4}$ and correct it
correspondingly. As shown in Table III, for the cases listed in the first
six lines, errors can be recognized and corrected. This identification
method can be used for the CV TEC but cannot be used for the qubit. For the
case of the last line in Table III, the syndrome measurements are the same
as those case where no error occurs. So it cannot be recognized. The
situations for the other ten possible cases for three identical\ errors
which are not listed in Table III are similar to the cases listed in Table
III. The total correction probability for the case of three errors is $%
2\times 9p^{3}(1-p)^{3}$.

Figure 4 shows the dependence of the error rate for the protected quantum
correlation on the single mode error rate $p$. The error rate of TEC for a
qubit is given by $%
P_{2}=1-p^{6}-(1-p)^{6}-6p^{5}(1-p)-6p(1-p)^{5}-9p^{4}(1-p)^{2}-9p^{2}(1-p)^{4}
$\cite{Yao2012}, which is shown by curve b in Fig. 4. It is obvious that the
error rate of TEC for a qubit (curve b) is reduced when compared with the
error rate $P_{1}$ without TEC (curve a). The error rate of our\ CV TEC
scheme is $P_{3}=P_{2}-2\times 9p^{3}(1-p)^{3},$ which is shown in Fig. 4.
It is obvious that the error rate of our CV TEC scheme is further reduced
than that of TEC for a qubit.

\section{DISCUSSION\ AND\ CONCLUSION}

In the proposed CV TEC, the local measurements of the optical modes can be
used in QC and TEC at same time. It is different from the case in the CV QEC
with a nine-wave-packet code \cite{Aoki2009} and a five-wave-packet code
\cite{Hao}, where the ancillary modes are measured to identify the position
of error in the error syndrome process and to remove the displacement error
on an input mode. The finite squeezing of the ancillary modes will introduce
extra noise in these CV QEC. However, in the proposed CV TEC, the
topological quantum correlations are protected against phase displacement
error instead of an optical mode. All optical modes involved in topological
quantum correlations are measured to identify the error and correct the
error. More interestingly, the error can be removed without extra noise
introduced in the case of finite squeezing.

Although the beam-splitter network for preparation of the eight-partite
topological\ cluster state seems a little bit complex, it can be easily
obtained by using photonic circuits. For example, a more complex photonic
circuit with a high quality of stability, matrix randomness and ultra-low
transmission loss has been used in a boson sampling experiment \cite{WangHui}%
. It is convenient to use this technique in our scheme.

The cluster state used in our scheme is similar to the \textit{N}-mode
multiple-rail cluster state introduced by P. van Loock \textit{et al}. in
Ref. \cite{vanLoock2007}, which is useful for error filtration in a Gaussian
cluster computation. The difference between the presented TEC scheme and
that of Ref. \cite{vanLoock2007} is as follows. First, as shown in Ref. \cite%
{vanLoock2007}, the multiple-rail cluster state has the possibility of a
noise reduction for the excess noise in a phase quadrature, which is the
noise added to the input state in the quantum teleportation protocol and
only depends on the nullifiers of the imperfect ancillary cluster state.
However, in our scheme, the topological quantum correlation in an
eight-partite CV cluster state is investigated, as shown in Eq. (5), and the
TEC scheme for protecting the topological quantum correlation against a
single quadrature phase displacement error occurring on any modes is
presented. Second, an input state is involved in the quantum teleportation
in Ref. \cite{vanLoock2007}. While in our scheme, no input state is
involved, only the tolerance for noise on the topological cluster state
itself is investigated.

Recently, it has been shown that Raussendorf-Harrington-Goyal (RHG) lattice
code is a very good candidate for fault-tolerant CV QC and it shows
robustness against analog errors during topologically protected
measurement-based QC \cite{Bourassa,Fukui}. However, in this paper, we
propose a CV TEC scheme to protect topological quantum correlation, which is
the CV analog of the TEC with an eight-photon graph state \cite{Yao2012} and
different from the RHG code. How to associate the presented TEC scheme with
the measurement-based CV QC remains an open question. It is worthwhile to
investigate the possibility of measurement-based QC with the presented TEC
scheme in the future.

In summary, the preparation scheme for a topological eight-partite CV
cluster state is proposed. Based on this special cluster state, we show that
topological quantum correlation can be protected against a single phase
quadrature displacement error occurring on any modes. Some cases of two
identical\ errors and three identical\ errors occurring simultaneously can
also be recognized and corrected. The details of recognition, correction and
the error rate of the CV TEC scheme are presented. We also show that the
error rate of the presented CV TEC scheme is lower than that of the TEC for
a qubit. The presented scheme has potential application in CV TEC and it is
feasible with current technology.

\section*{ACKNOWLEDGMENTS}

This research was supported by the NSFC (Grants No. 11804001, No. 11834010,
No. 61675006), the Natural Science Foundation of Anhui Province (Grant Nos
1808085QA11), the program of Youth Sanjin Scholar, National Key R\&D Program
of China (Grant No. 2016YFA0301402), and the Fund for Shanxi
\textquotedblleft 1331 Project\textquotedblright\ Key Subjects Construction.

\end{document}